\documentclass{aa}[referee]
\usepackage[english]{babel}
\usepackage{microtype}
\usepackage{txfonts}

\usepackage{mathtools}
\usepackage{bm}
\usepackage{mathrsfs}
\usepackage{nccmath}

\usepackage[debug]{hyperref}
\hypersetup{colorlinks=true, allcolors=blue}

\usepackage{graphicx}

\makeatletter
\DeclareRobustCommand\onedot{\futurelet\@let@token\@onedot}
\def\@onedot{\ifx\@let@token.\else.\null\fi\xspace}
\def\eg{\textit{e.g}\onedot}
\def\ie{\textit{i.e}\onedot}

\makeatother

\renewcommand{\vec}[1]{\bm{#1}}
\newcommand{\1}{{\mathbf{1}}}
\newcommand{\0}{{\mathbf{0}}}

\newcommand{\grad}{\vec{\nabla}}

\newcommand{\Dt}{{\mathscr{D}_t}}
\renewcommand{\dot}[1] {\overset{\,_{\mbox{\Large .}}}{#1}}

\begin{document}

\title{
    Phase-space averaging for stellar convection
}
\subtitle{
    I. Liouvillian dynamics
}

\author{
	P.~S.~Houdayer \email{pierre.houdayer@utoulouse.fr}
	\and 
	M.~Rieutord \email{michel.rieutord@utoulouse.fr}
}

\institute{
	Univ Toulouse, CNRS, CNES, IRAP, 14, avenue Édouard Belin, F-31400 Toulouse, France.
}

\date{Received 28 May 2026; Accepted 23 July 2026}

\abstract
{
Convection remains one of the main uncertain links between multidimensional hydrodynamics and one-dimensional stellar evolution.
In particular, transition regions such as near-surface layers or convective boundaries require mean-field descriptions that remain connected to the underlying dynamics rather than to a prescribed mixing length.
}
{
We aim to build a dynamical mean-field framework for stellar convection in which the averaging procedure, the hierarchy of moments, and the local stability diagnostic all follow from the motion of convective fluid elements.
This provides the theoretical basis for the closure model developed in the companion paper.
}
{
We describe the flow as a distribution of mesoscopic fluid particles in position--velocity space.
A conservation law for this distribution defines the average and yields the Reynolds-Favre mean-field equations as velocity-space moments.
Under standard interior conditions, the same dynamics can be written in Liouvillian form, which extends the Hamiltonian structure to stratified and dissipative media.
}
{
The Liouvillian formulation identifies the phase-space divergence, \(\{s, T\}\), as a local measure of contraction or expansion of nearby trajectories.
In the quasi-adiabatic limit, its sign reduces to the classical Schwarzschild stability criterion.
Away from this limit, the diagnostic remains velocity-resolved and can distinguish different parts of the convective population within the same layer, for example in surface and penetration regions.
Appendices show how rotation, magnetic fields, and composition changes can be incorporated through modifications of the phase-space structure.
}
{
Phase-space averaging provides a dynamically grounded route from hydrodynamics to mean-field stellar convection equations.
It also supplies a local trajectory-based stability diagnostic and a natural starting point for the maximum-entropy closures constructed in the companion paper.
}


\maketitle
\nolinenumbers

\section{Introduction}
\label{sec_intro}

From stellar cores to photospheres, thermal convection links local fluid dynamics to the global structure of stars. 
It governs the redistribution of energy and chemicals over a wide range of scales and thus shapes both the internal and observable properties of stars. 
In stellar cores, convection sets the efficiency of nuclear burning \citep{Maeder1987}, controls the supply of fresh fuel by overshooting \citep{Maeder1975, Roxburgh1989, Zahn1991, Canuto1997b}, and thereby affects stellar ages and luminosities \citep{Maeder1989, Chiosi1992, Deheuvels2016}. 
In stellar envelopes, it regulates energy transfer in the outer layers \citep{Cattaneo1991, Nordlund2009, Meakin2010, Arnett2015}, excites acoustic and gravity waves \citep{Goldreich1977, Goldreich1990, Samadi2001a}, determines the extent of convective regions \citep{Schwarzschild1958, Deng2006, Arnett2015}, sustains surface magnetic activity \citep{Cattaneo2003, Stein2012}, and drives penetration and mixing at convective boundaries \citep{Zahn1991, Brummell2002, Rempel2004}. 
Capturing these phenomena in stellar models requires a description of convection that connects turbulent, three-dimensional flows to the quasi-stationary, one-dimensional structure used in evolution calculations.

In practice, thermal convection in one-dimensional stellar evolution codes is still treated with mixing-length theory (MLT) and its numerous variants \citep{Opik1950, BohmVitense1958, Unno1967}. 
Despite its simplicity, MLT has proved remarkably robust: it enables full stellar-structure computations, admits a solar calibration, and often reproduces global trends in HR diagrams at modest computational cost \citep[see][for a recent review]{Joyce2023}. 
At the same time, it rests on a phenomenological picture of a representative convective element travelling over a prescribed mixing length. 
The link between this picture and the underlying hydrodynamic equations is indirect: no explicit averaging operator connects the local, three-dimensional flow to the one-dimensional mean fields, and essential mean quantities such as convective velocities, entropies, and fluxes lack a unique statistical or dynamical interpretation.

Many extensions have been proposed to improve specific aspects of MLT, including non-local \citep[e.g.][]{Spiegel1963, Ulrich1970b, Balmforth1992a, Deng2006}, time-dependent \citep{Schatzman1956, Unno1967, Gough1977, Gabriel1996, Grigahcene2005}, and multi-scale or Reynolds-stress approaches \citep{Canuto1991, Canuto1996, Xiong2021}. 
In parallel, three-dimensional radiation-hydrodynamics simulations have become a key tool to study near-surface convection and to calibrate one-dimensional models \citep{Nordlund2009, Freytag2012, Kupka2017}. 
Across this landscape, an enduring practical difficulty is to define mean fields and closures in a way that remains tightly connected to the underlying equations of motion, particularly in stratified, radiatively coupled, and transitional layers.

The present work introduces a phase-space averaging (PSA) formalism in which the statistical description is built directly from the hydrodynamic motion of convective fluid elements.
The objective is to derive both the mean-field equations and a local stability diagnostic from a single phase-space dynamics, rather than introducing the averaging procedure and the stability criterion as separate ingredients.
This provides a dynamical foundation for the closure strategy developed in the companion paper.

The paper follows this construction step by step.
Section~\ref{sec:psa} defines the PSA framework and derives the Reynolds-Favre mean-field equations as moments of the velocity distribution.
Section~\ref{sec:liouvillian} develops the Liouvillian formulation, identifies the phase-space divergence \(\{s,T\}\), and shows how it connects to the Schwarzschild criterion in the quasi-adiabatic limit.
Section~\ref{sec:discussion} discusses the assumptions, limitations, and possible extensions of the formalism, before the conclusions are given in Sect.~\ref{sec:conclusion}.
The companion paper then uses this framework to construct entropy-maximising closures and a practical model for one-dimensional convective envelopes.

\section{Phase-space averaging formalism}
\label{sec:psa}

\subsection{Classical averaging and its limitations}

Mean-field descriptions of convection generally rely on separating fields into mean and fluctuating parts.  
For a quantity defined per unit volume, \(y\), the Reynolds average \(\bar{y}\) is introduced as
\begin{equation}
    \label{eq:Reynolds_average}
    y = \bar{y} + y', \qquad \bar{y'} = 0,
\end{equation}
while for a variable defined per unit mass, \(x\), the Favre average \citep{Favre1965} is preferred,
\begin{equation}
    \label{eq:Favre_average}
    \tilde{x} = \frac{\overline{\rho x}}{\bar{\rho}}, \qquad 
    x = \tilde{x} + x'', \qquad 
    \overline{\rho x''} = 0.
\end{equation}
These definitions are convenient when manipulating the equations of hydrodynamics, as they allow one to write the conservation laws in terms of mean and fluctuating components.  
They also ensure that products such as \(\overline{\rho x}\) or \(\overline{y_1y_2}\) can be decomposed consistently into mean and covariance terms.

A common working assumption in mean-field manipulations is that the averaging operator commutes with temporal and spatial derivatives,
\begin{equation}
    \overline{\partial_t y} = \partial_t \bar{y}, \qquad
    \overline{\grad y} = \grad \bar{y}.
\end{equation}
This property is exact only for some idealised choices (for instance, stationary time averages with good scale separation), and it is otherwise adopted as a practical approximation.
In stellar convection, however, strong density stratification, radiative coupling, and the lack of homogeneity make it natural to seek an averaging operator that is tied more explicitly to the local dynamics rather than imposed a priori.

A useful guide is provided by the kinetic theory of gases, where macroscopic fields are obtained as moments of a phase-space distribution \(f(t,\vec q,\vec p)\), and the closure of hydrodynamics relies on an explicit separation between microscopic and macroscopic scales \citep{Chapman1970}.
In stellar interiors, the separation between microscopic scales (mean free paths and collision times) and global stellar scales is enormous, leaving a wide mesoscopic range in which continuum thermodynamics already applies while the flow remains far from the global, one-dimensional mean fields used in stellar models \citep[e.g.][]{Kupka2017}.
It is within this intermediate range that we place our description.

\subsection{Mesoscopic phase-space description}

\begin{figure*}
    \centering
    \includegraphics[width=\textwidth]{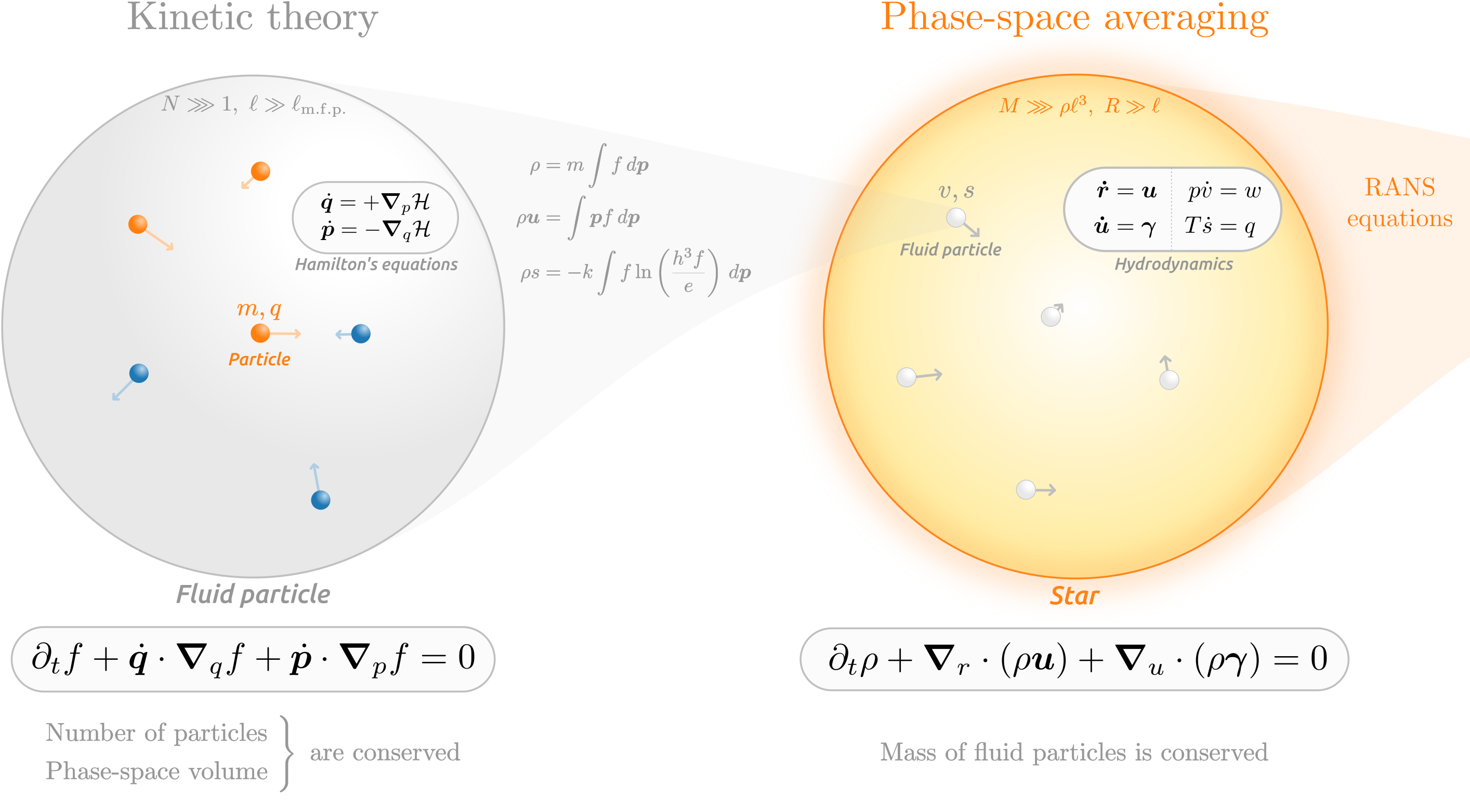}
	\caption{
		Schematic analogy between kinetic theory and phase-space averaging (PSA).
		\textit{Left:} In kinetic theory, microscopic particles populate \((\vec q,\vec p)\) and are described by a distribution \(f(t,\vec q,\vec p)\) transported by Liouville’s equation under Hamilton’s equations, so the phase-space flow is incompressible.
		Velocity moments of \(f\) yield macroscopic fields once a separation of scales exists.
		\textit{Right:} In PSA, the star is described in terms of mesoscopic fluid particles of size \(\ell\), with \(\ell \gg \ell_{\rm m.f.p.}\), carrying thermodynamic variables \((v, s)\) and kinematics \((\vec r,\vec u)\) evolving through the hydrodynamics.
		Their phase-space mass density \(\rho(t,\vec r,\vec u)\) satisfies the continuity equation shown at the bottom.
	}
    \label{fig:psa_hierarchy}
\end{figure*}

Classical mean-field descriptions are often introduced through averaging procedures over time, space, or ensembles of possible realisations.
While such constructions are convenient, the averaging operator is not, in general, defined by the dynamics of the flow itself.
Here we instead introduce a distribution function defined directly from the mesoscopic motion of the fluid in phase space.

We consider an ensemble of mesoscopic fluid elements, referred to hereafter as \textit{fluid particles}, and described by a mass density in phase space \(\rho(t,\vec r,\vec u)\).
Each fluid particle is large enough that microscopic kinetic degrees of freedom have relaxed to local thermodynamic equilibrium, so that quantities per unit mass such as specific entropy \(s\), specific volume \(v\), and temperature \(T\) are well defined, yet small enough compared with stellar-structure scales.

By construction, \(\rho\) is a mass density on phase space: for any element \(d\vec r\,d\vec u\), the quantity \(dM = \rho(t,\vec r,\vec u)\,d\vec r\,d\vec u\) represents the mass of fluid particles located near position \(\vec r\) and moving with velocity \(\vec u\) at time \(t\).
The usual macroscopic density field \(\bar{\rho}(\vec r,t)\) is recovered as the marginal of \(\rho\) over velocity space,
\begin{equation}
    \label{eq:rho_marginal_new}
    \int \rho(t,\vec r,\vec u)\,d\vec u = \bar{\rho}(\vec r,t),
\end{equation}
where the integral over \(\vec u\) is understood with a fixed measure on velocity space which is defined in the next subsection.

The analogy with kinetic theory \citep{Chapman1970} provides a useful guide.  
There, the collective behaviour of a very large number of microscopic particles---atoms or molecules---is described by a distribution function \(f(t,\vec q,\vec p)\) defined on the microscopic phase space \((\vec q,\vec p)\) and normalised to the total number of particles,
\begin{equation}
    \int f(t, \vec{q}, \vec{p})\,d\vec p = n(\vec{q},t),
\end{equation}
where \(n(\vec q,t)\) is the local number density.
The trajectories of individual particles, \(\vec \varphi(t) = (\vec q(t),\vec p(t))\), obey Hamilton’s equations,
\begin{equation}
    \dot{\vec q} = \grad_{p} H, 
    \qquad
    \dot{\vec p} = -\,\grad_{q} H,
\end{equation}
so that the associated flow in phase space is incompressible:
\begin{equation}
    \label{eq:Liouville_theorem_new}
    \operatorname{div}\,(\dot{\vec \varphi}) 
    = (\grad_{q} \cdot \grad_{p} - \grad_{p} \cdot \grad_{q}) H = 0,
\end{equation}
which expresses Liouville’s theorem, i.e. the conservation of phase-space volume along Hamiltonian trajectories.
As a result, the microscopic distribution \(f\) is transported by a conservative Liouville equation.

In the present context, the correspondence is similar in spirit but differs in physical content.  
We start from hydrodynamics and define the trajectories of fluid particles in the mesoscopic phase space \((\vec r,\vec u)\) through
\begin{equation}
    \dot{\vec r} = \vec u, \qquad 
    \dot{\vec u} = \vec \gamma ,
\end{equation}
where \(\vec \gamma \) is the acceleration arising from local hydrodynamic forces such as pressure gradients and gravity.  
The evolution of the distribution \(\rho\) then follows a continuity equation in the six-dimensional space \((\vec r,\vec u)\),
\begin{equation}
    \label{eq:continuity_phase_space}
    \partial_t \rho
    + \grad_{r} \cdot (\rho \vec u)
    + \grad_{u} \cdot (\rho \vec \gamma ) = 0.
\end{equation}
Equation~\eqref{eq:continuity_phase_space} ensures conservation of total mass in phase space and is the analogue, at the fluid-particle level, of the Liouville equation in kinetic theory.

However, unlike the Hamiltonian case, there is no guarantee that the phase-space flow generated by \((\dot{\vec r},\dot{\vec u})\) is incompressible. 
The hydrodynamic velocity and acceleration do not in general derive from a common potential: dissipative processes are present, and the associated flow in \((\vec r,\vec u)\) need not conserve phase-space volume.  
For this reason, Eq.~\eqref{eq:continuity_phase_space} is taken here as a primary statement of mass conservation, without assuming phase-space incompressibility.

This dynamical formulation provides a phase-space alternative to abstract stochastic-ensemble constructions.
In this description, each trajectory \((\vec r(t),\vec u(t))\) represents the motion of a fluid particle, and the statistical properties of the flow are encoded directly in the distribution \(\rho(t,\vec r,\vec u)\).

\subsection{Mean-field quantities as phase-space moments}
\label{sec:moments}

The phase-space distribution \(\rho(t,\vec r,\vec u)\) allows any quantity \(x(t,\vec r,\vec u)\) attached to a fluid particle to be averaged directly over velocity space.
We introduce the velocity-moment operator
\begin{equation}
    \label{eq:integral_u}
    \langle \bullet \rangle(\vec r,t) \equiv \int \bullet \, d\vec u,
\end{equation}
where \(d\vec u\) is the (fixed) Lebesgue measure on velocity space and all velocity integrals are assumed convergent under the decay of \(\rho\) as \(|\vec u|\to\infty\).
The phase-space mass density \(\rho\) defines the mean density field as its zeroth moment,
\begin{equation}
    \bar\rho(\vec r,t)=\langle \rho\rangle,
\end{equation}
and induces the conditional velocity distribution
\begin{equation}
    f(\vec u \mid \vec r, t)\equiv \frac{\rho(t, \vec r,\vec u)}{\bar\rho(\vec r, t)}, \qquad \langle f\rangle = 1.
\end{equation}
The Favre mean of any specific quantity \(x(t, \vec r,\vec u)\) is then the expectation under \(f\),
\begin{equation}
    \label{eq:PSA_average}
    \tilde{x}(\vec r,t)=\frac{\langle \rho x\rangle}{\bar\rho}=\langle f x\rangle,
\end{equation}
and we define the associated Favre fluctuation by \(x''\equiv x-\tilde x\), so that \(\langle \rho x''\rangle=0\).
For volumetric quantities \(y\equiv\rho x\), the corresponding mean field is simply the velocity moment
\begin{equation}
    \bar y(\vec r,t)\equiv \langle y\rangle=\langle \rho x\rangle=\bar\rho \tilde x,
\end{equation}
which provides the PSA counterpart of Reynolds averaging.
A key advantage of this construction is that the commutation with derivatives (which is often assumed, or only approximately justified, in classical mean-field manipulations) holds here exactly, because the averaging is performed at fixed \((\vec r,t)\) with a fixed velocity measure, hence:
\[
    \bigl \langle \partial_t(\rho x) \bigr \rangle
    = \partial_t \langle \rho x \rangle, \qquad
    \bigl \langle \grad_{r} (\rho x) \bigr \rangle
    = \grad_{r} \langle \rho x \rangle .
\]

Let \(x\) be any scalar quantity that satisfies an evolution law of the form
\begin{equation}
    \label{eq:x_evolution_general}
    \Dt x = \dot{x},
\end{equation}
where \(\Dt = \partial_t + \vec u \cdot \grad_{r} + \vec \gamma  \cdot \grad_{u}\) denotes the material derivative in \((\vec r,\vec u)\) and \(\dot{x}\) the variation of \(x\) along the trajectory.
Denoting by \(\vec{Q}_x, \vec{R}_x\) and \(S_x\) the physical and velocity space fluxes and source term per unit mass associated with \(x\) at the scale of an individual fluid particle, the local conservation law for \(x\) reads
\begin{equation}
    \rho \dot{x} = - \grad_{r} \cdot \vec{Q}_x - \grad_{u} \cdot \vec{R}_x + \rho S_x.
\end{equation}

Multiplying Eq.~\eqref{eq:continuity_phase_space} by \(x\) and using Eq.~\eqref{eq:x_evolution_general} one obtains
\begin{equation}
    \partial_t (\rho x)
    + \grad_{r} \cdot (\rho x \vec u)
    + \grad_{u} \cdot (\rho x \vec \gamma ) = \rho \dot{x}.
\end{equation}
Using properties \eqref{eq:Favre_average} and \eqref{eq:integral_u} and integrating over velocity space yields, assuming boundary terms in velocity space vanish,
\begin{equation}
    \label{eq:moment_general}
    \partial_t \langle \rho x \rangle
    + \grad \cdot
      \Big[\langle \rho x \rangle\,\tilde{\vec u}
            + \langle \rho x''\vec u''\rangle
            + \langle \vec{Q}_x \rangle \Big]
    = \langle \rho S_x \rangle .
\end{equation}

Equation~\eqref{eq:moment_general} gives the general form of the averaged conservation law for any quantity \(x\).
The term \(\langle \rho x''\vec u''\rangle\) identifies with the flux of turbulent correlations in the classical mean-field formalism, while \(\langle \vec{Q}_x \rangle\) and \(\langle \rho S_x \rangle\) account for the averaged transport and source contributions.
Velocity-space fluxes such as \(\vec{R}_x\) enter the local conservation law only through \(\grad_{u} \cdot \vec{R}_x\); after integration over \(\vec u\), these contributions reduce to surface terms in velocity space and vanish under the usual decay assumptions at \(|\vec u|\to\infty\), so they do not appear explicitly in the mean-field equations.

The first three moments of the continuity equation, corresponding respectively to \(x = 1\), \(x = \vec u\), and \(x = \vec u^{\,2}/2\), reproduce the mean-field equations of mass, momentum, and kinetic-energy conservation:
\begin{align}
    \label{eq:mass_conservation}
    &\partial_t \bar{\rho} + \grad \cdot \langle \rho \vec u \rangle = 0, \\[3pt]
    \label{eq:momentum_conservation}
    &\partial_t \langle \rho \vec u \rangle
        + \grad \cdot \Big(\langle \rho \vec u \rangle \,  \tilde{\vec u} + \mathrm R\Big)
        = \langle \rho \vec \gamma  \rangle, \\[3pt]
    \label{eq:kinetic_energy_conservation}
    &\partial_t \langle \rho k \rangle
        + \grad \cdot \Big( \langle \rho k \rangle \, \tilde{\vec u}
        + \vec{Q}_k \Big)
        = \langle \rho \vec u \cdot \vec \gamma  \rangle,
\end{align}
with \(k = \vec u^{\,2}/2\) the kinetic energy and
\begin{equation}
    \mathrm R = \langle \rho\,\vec u''\vec u''\rangle,
    \qquad
    \vec{Q}_k = \langle \rho k''\vec u''\rangle
\end{equation} 
respectively denoting the Reynolds stress tensor and the kinetic-energy flux.
Equations~\eqref{eq:mass_conservation}-\eqref{eq:kinetic_energy_conservation} are formally identical to the Favre-averaged hydrodynamic equations.

Since quantities such as internal energy or entropy are defined at the level of individual fluid particles, the same procedure can be applied to derive the transport equation for the mean specific enthalpy \(\tilde{h}\):
\begin{equation}
    \label{eq:enthalpy_conservation}
    \partial_t \langle \rho h \rangle
    + \grad \cdot
      \Big[\langle \rho h \rangle\,\tilde{\vec u}
            + \langle \rho h''\vec u''\rangle
            + \langle \vec{Q}_h \rangle \Big]
    = \langle \rho S_h \rangle,
\end{equation}
where \(\vec{Q}_h\) and \(S_h\) denote, respectively, the enthalpy flux and source per unit mass at the fluid-particle scale. In the stellar context, these terms represent radiative (and possibly conductive) transport and the local production of enthalpy by viscous dissipation, compressibility, or nuclear reactions \citep{Canuto1997}.
Equations~\eqref{eq:kinetic_energy_conservation} and \eqref{eq:enthalpy_conservation} can then be combined to yield the mean-field transfer of total energy, confirming that the mean-field equations emerge naturally as velocity-space moments of the phase-space continuity law. The PSA framework therefore provides a unified and dynamically grounded derivation of the mean-field equations, in which the Reynolds and Favre averages appear as direct consequences of phase-space integration.

\section{Liouvillian dynamics and stability}
\label{sec:liouvillian}

\subsection{Entropy balance in phase space}
\label{sec:entropy_balance}

A natural question, in view of the analogy with kinetic theory, is whether a Liouville-type property applies again at the level of fluid particles.  
Starting from the phase-space continuity law, Eq.~\eqref{eq:continuity_phase_space}, one readily obtains the kinematic identity
\begin{equation}
    \label{eq:entropy_phase_space}
    \operatorname{div} \, (\dot{\vec \varphi}) = \Dt(-\ln \rho),
\end{equation}
where \(\vec \varphi(t) = (\vec r(t),\vec u(t))\) denotes the fluid-particle position in phase space.
Thus, conservation of phase-space volume along trajectories, \(\operatorname{div} \, (\dot{\vec \varphi}) = 0\), is equivalent to the conservation of the local specific statistical entropy \(-\ln\rho\) along those trajectories.

While this statement is familiar for the joint \(N\)-particle distribution in kinetic theory, it is lost at the level of the one-particle distribution, where inter-particle interactions must be accounted for explicitly \citep{Chapman1970}.
For fluid particles, collective interactions take the form of heat and mechanical work exchange, and there is likewise no reason to expect incompressibility in \((\vec r,\vec u)\).
Quantitatively, an ``adiabatic'' displacement in phase space (in the sense that it would preserve the entropy of the distribution \(\rho\)) would amount to
\begin{equation}
    \operatorname{div} \, (\dot{\vec \varphi}) = \grad_{r} \cdot \vec u + \grad_{u} \cdot \vec \gamma = 0,
\end{equation}
which requires an exact compensation between dilatation/compression in real space and in velocity space.  
Such a cancellation has, \textit{a priori}, no reason to hold in the stellar context.

\subsection{Liouvillian generator for fluid-particle dynamics}
\label{sec:liouvillian_generator}

Having established that the phase-space flow need not be incompressible, we now ask whether its dynamics can nevertheless be generated by a potential on \((\vec r,\vec u)\), in the same spirit as Hamiltonian flows.
In kinetic theory, this structure underlies Liouville’s theorem; here we will see that an analogous construction is possible and, crucially, that it cleanly separates conservative and non-conservative contributions.
Two simplifications are however needed for this construction:
\begin{enumerate}[(i)]
	\item \textit{immediate pressure adjustment}, so that pressure does not depend on the velocity coordinate,
	\[
	    \grad_{u} p = \vec 0,
	\]
	which is expected to fail as soon as motions become supersonic; \newline
	\item \textit{large Reynolds number}, so that viscous stresses are negligible and the fluid-particle acceleration reads
	\[
	    \vec \gamma = \vec g - v \grad_{r} p,
	\]
	with \(v\) the specific volume and \(\vec g = -\grad_{r}\phi\) the gravitational acceleration.
\end{enumerate}
Let us now introduce the fluid-particle Hamiltonian
\begin{equation}
    H(\vec r,\vec u)
    = \varepsilon(\vec r,\vec u)
      + p(\vec r)\,v(\vec r,\vec u)
      + \phi(\vec r)
      + \frac12\,\vec u^{\,2},
\end{equation}
which collects the internal energy \(\varepsilon\), the thermodynamic potential \(pv\) (work against pressure per unit mass), the gravitational potential \(\phi\), and the kinetic energy \(\vec u^{\,2}/2\) of a fluid particle. 
We make the dependencies of \(H\) explicit for clarity below. Using the first law of thermodynamics, \(d \varepsilon = T ds - p dv\), one obtains
\begin{align}
    \grad_{r} H &= T\,\grad_{r} s + v\,\grad_{r} p + \grad_{r}\phi
                             = T\,\grad_{r} s - \vec \gamma , \\
    \grad_{u} H &= T\,\grad_{u} s + \vec u .
\end{align}

Given these relations, it is natural to introduce the \textit{Liouvillian}\footnote{
	We use the term \emph{Liouvillian} in the standard sense of statistical mechanics, namely as the generator of the phase-space transport equation for \(\rho\).
	Unlike in the Hamiltonian case, the associated phase-space flow need not be incompressible, so \(\psi\) denotes here a non-Hamiltonian generator appropriate to dissipative media; in practice, only its differential \(d\psi\) is used below, and \(\psi\) itself need not be globally defined.
} \textit{generator} through
\begin{equation}
    d \psi = d H - Td s .
\end{equation}
In other words, \(d\psi\) removes from \(H\) the explicitly non-conservative contribution associated with entropy production.
The equations of motion in \((\vec r,\vec u)\) then follow as
\begin{equation}
    \label{eq:Liouville_motion}
    \dot{\vec r} = \grad_{u}\psi, 
    \qquad
    \dot{\vec u} = - \grad_{r}\psi,
\end{equation}
which generalise Hamilton’s equations to dissipative motions (\(ds \neq 0\)). 
Accordingly, for any quantity \(x(t,\vec r,\vec u)\),
\begin{equation}
    \Dt x = \partial_t x + \{x, \psi\},
    \qquad
    \{x,\psi\} \equiv \grad_{u} \psi \cdot \grad_{r} x - \grad_{r} \psi \cdot \grad_{u} x .
\end{equation}
This Poisson bracket on \((\vec r,\vec u)\) mirrors the Hamiltonian case while consistently incorporating thermodynamics through \(\psi\), which acts as the generator of the fluid-particle dynamics in phase space.

\subsection{Phase-space divergence and stability interpretation}
\label{sec:phase_space_divergence}

\begin{figure*}
    \centering
    \includegraphics[width=\textwidth-1.5cm]{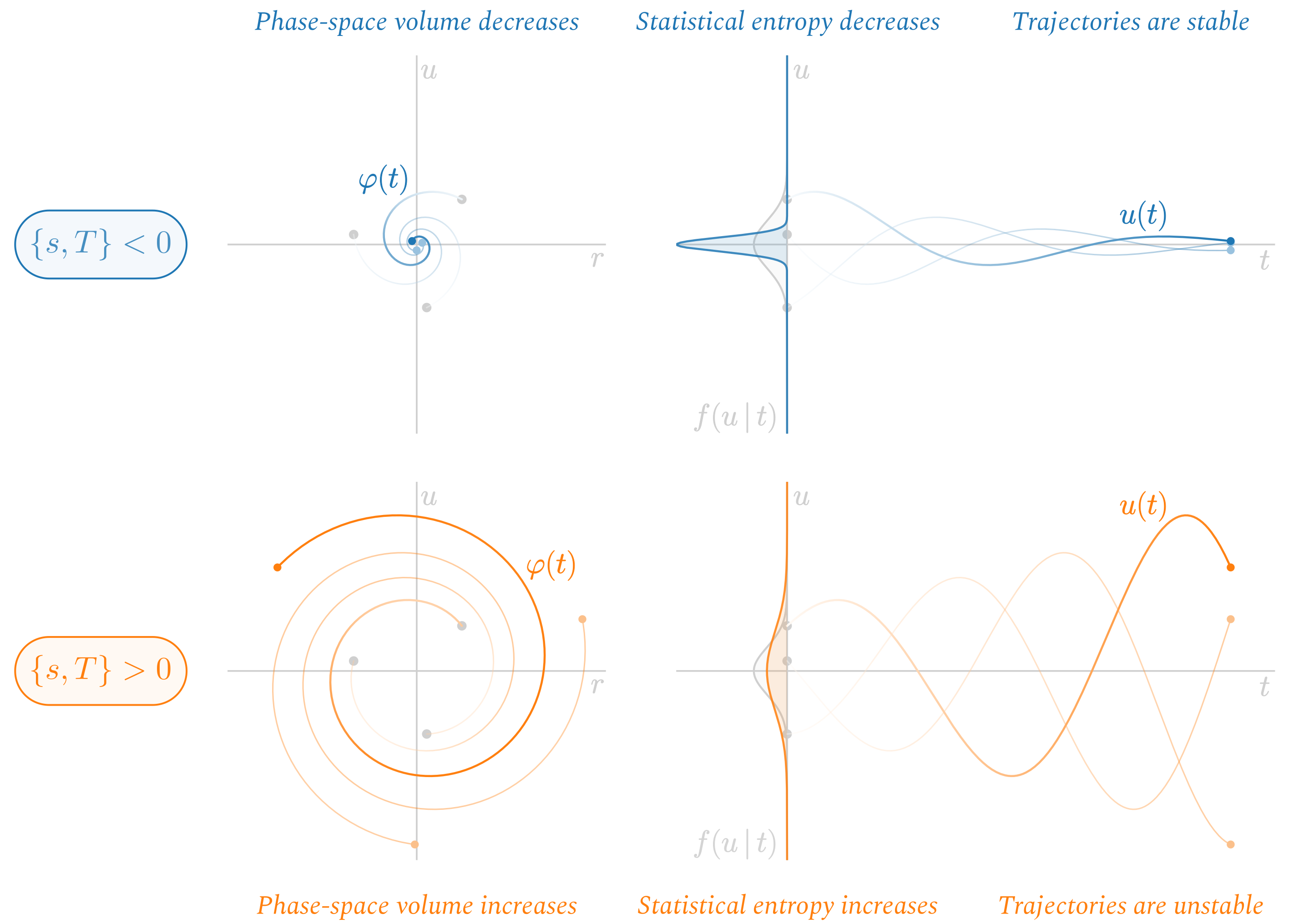}
	\caption{
		Schematic interpretation of the phase-space divergence \(\operatorname{div} \, (\dot{\vec \varphi})=\{s,T\}\) in a two-dimensional \((r,u)\) slice.
		In each row, the left panel illustrates the deformation of a small set of initial conditions in \((r,u)\), while the right panel sketches the associated evolution of \(u(t)\) and the corresponding concentration or dilution of the realised velocity distribution.
	}
    \label{fig:phase_space_stability}
\end{figure*}

Starting from the Liouvillian equations of motion \eqref{eq:Liouville_motion}, the divergence of the phase-space flow reads
\begin{equation}
    \label{eq:div_flow_psi}
    \operatorname{div} \, (\dot{\vec \varphi})
    = \grad_{r} \cdot(\grad_{u} \psi) - \grad_{u} \cdot(\grad_{r} \psi)
    = \big(\grad_{r} \cdot \grad_{u} - \grad_{u} \cdot \grad_{r} \big)\psi .
\end{equation}
Because \(H\) is a scalar potential, its mixed derivatives commute in Eq.~\eqref{eq:div_flow_psi}.
Hence the conservative part \(dH\) cancels, and only the non-conservative thermal contribution \(-T ds\) can generate a non-zero divergence in \((\vec r,\vec u)\).
A short calculation gives
\begin{equation}
    \label{eq:div_as_bracket}
    \operatorname{div} \, (\dot{\vec \varphi})
    = -\, \big(\grad_{r} T \cdot \grad_{u} s - \grad_{u} T \cdot \grad_{r} s\big)
    = \{s,T\},
\end{equation}
that is, the divergence equals the Poisson bracket between entropy and temperature.

Equation~\eqref{eq:div_as_bracket} provides a closed, purely kinematic expression for the phase-space divergence (compressibility) of the Liouvillian flow in phase space.
To make its meaning explicit, consider the flow map \(\vec \varphi(t; \vec \varphi_0)\) generated by Eq.~\eqref{eq:Liouville_motion}, and let \(J(t)=\det \, (\partial \vec \varphi(t)/\partial \vec \varphi_0)\) be its Jacobian.
A standard identity for smooth flows gives
\begin{equation}
    \label{eq:jacobian_div}
    \frac{d}{dt}\ln J = \operatorname{div} \, (\dot{\vec \varphi}) = \{s,T\},
\end{equation}
so that an infinitesimal phase-space volume element evolves as
\begin{equation}
    \label{eq:volume_element}
    d\vec \varphi(t) =
    \exp\left(\int_0^t \{s,T\}\big(\vec \varphi(t')\big)\,dt'\right)  d\vec \varphi_0.
\end{equation}
In other words, \(\{s,T\}\) is the instantaneous rate of contraction (\(\{s,T\}<0\)) or expansion (\(\{s,T\}>0\)) of the Liouvillian flow in \((\vec r,\vec u)\).
Because \(\operatorname{div} \, (\dot{\vec \varphi})\) also satisfies the kinematic identity \eqref{eq:entropy_phase_space}, the same quantity controls the local concentration or dilution of the phase-space mass density along trajectories:
\begin{equation}
    \label{eq:rho_along_flow}
    \Dt\ln\rho = -\,\{s,T\}.
\end{equation}

Figure~\ref{fig:phase_space_stability} provides a schematic illustration of these implications in a two-dimensional slice \((r,u)\).
The left panels emphasise the deformation of a small set of initial conditions in phase space, while the right panels sketch how the associated velocity distribution may narrow or broaden as trajectories refocus \((\{s,T\}<0)\) or defocus \((\{s,T\}>0)\).
In that sense, the phase-space divergence \(\operatorname{div} \, (\dot{\vec \varphi})=\{s,T\}\) can be read as a local Lagrangian stability criterion: it measures the instantaneous logarithmic rate at which nearby fluid-particle trajectories collectively contract or expand in phase space.
This notion of stability is therefore trajectory-based rather than modal, and should not be confused with a normal-mode growth rate in the usual linear-stability sense.
It is then natural to ask under which conditions \(\{s,T\}(\vec r,\vec u)\) can be reduced to a purely positional criterion, and how it connects to the Schwarzschild criterion.

As shown in Appendix ~\ref{app:schwarzschild_limit}, in a regime where:
\begin{enumerate}[(i)]
	\item \textit{locality} holds over the acceleration time (the relevant excursions remain local with respect to the background stratification),
	\item \textit{quasi-adiabaticity} holds over that time (non-advective entropy exchange is negligible) and
	\item entropy fluctuations within the local velocity population are \textit{small},
\end{enumerate}
the phase-space entropy gradient \(\grad_{u} s\) becomes slaved to the spatial gradient of the mean entropy, \(\grad_{r} \tilde s\).
In this limit, the phase-space divergence criterion reduces to the usual layer-wise sign test, so that
\begin{equation}
    \label{eq:sign_match_schwarzschild}
    \mathrm{sign}\,\{s,T\} = \mathrm{sign}\,(-N^2),
    \qquad
    N^2 = - \, \frac{\vec g \cdot \grad_{r}\tilde s}{c_p},
\end{equation}
with \(N\) the Brunt-V\"ais\"al\"a frequency.
In the schematic representation of Fig.~\ref{fig:phase_space_stability}, this correspondence amounts to associating the refocusing/defocusing behaviour with the stable/unstable nature of the local stratification.

Equation~\eqref{eq:sign_match_schwarzschild} therefore provides a sign-consistency check with the classical Schwarzschild indicator in the local quasi-adiabatic regime.
In practice, this reduction is useful only if assumptions (i)--(iii) hold over the velocities that effectively populate the distribution at fixed position, \ie over the range of \(\vec u\) where \(\rho(t,\vec r,\vec u)\) is non-negligible.
When this is the case, \(\{s,T\}\) has essentially a uniform sign across the realised velocity population and a single layer-wise label remains meaningful.
Otherwise, \(\{s,T\}\) must be interpreted as a genuinely phase-space diagnostic, and different velocity classes within the same layer may carry different signs.
We return to the implications of this distinction for radiative, convective, and transitional stellar layers in Sect.~\ref{sec:discussion_stellar_layers}.

\section{Discussion}
\label{sec:discussion}

\subsection{Assumptions, domain of validity, and limitations}
\label{sec:validity}

The phase-space description developed in Sects.~\ref{sec:psa}-\ref{sec:liouvillian} relies on a small set of structural assumptions that delimit its range of validity and clarify which physical effects are retained or excluded. 
We briefly summarise them here, and indicate the regimes in which the Liouvillian construction used in this paper is expected to require extension.

\subsubsection{Mesoscale fluid particles and local thermodynamic equilibrium.}

We assume the fluid to be strongly collisional on the scales of interest, so that a single temperature and a single scalar pressure field can be defined and shared by all species. 
This implies a genuine mesoscale window between microscopic kinetic scales and the macroscopic stellar stratification. 
Fluid particles must be large enough compared with the mean free path for local thermodynamic equilibrium to hold, yet small enough compared with the pressure scale height and the depth of the convective region for a non-trivial velocity distribution at fixed \((\vec r,t)\) to be meaningful. 
In kinetic-theory language, \(\rho(t,\vec r,\vec u)\) describes coarse-grained ``blobs'' whose internal degrees of freedom have thermalised, but which remain numerous enough for a distribution in \((\vec r,\vec u)\) to be defined. 
This separation is comfortably satisfied in most stellar interiors, but it progressively erodes as the radiative mean free path grows in the outermost layers, where the notion of a fluid-particle ensemble becomes less accurate.

\subsubsection{High Reynolds number and low Mach number.}

We work in a regime where viscous stresses do not contribute at leading order to the acceleration of fluid particles, so that the dynamics is dominated by pressure gradients and gravity. 
Non-conservative effects then enter primarily through radiative exchange and microphysical entropy production, encoded in the thermal part of the Liouvillian differential. 
In addition, the construction is tailored to low-Mach-number flows, where pressure remains close to a quasi-hydrostatic field enslaved to the thermodynamic state. 
This is the regime relevant to deep stellar interiors and to much of the subsonic envelope convection encountered in stellar evolution modelling.

\subsubsection{Limitations: transonic motions and dynamical pressure.}

The present Liouvillian construction is not limited by strong stratification or large thermodynamic contrasts per se, but by regimes where the flow approaches the transonic range. 
When \({\rm Ma}\sim 1\), pressure fluctuations acquire their own dynamics and are no longer enslaved to the instantaneous thermodynamic state, so the assumption \(\nabla_u p=\vec 0\) breaks down and the acceleration cannot, in general, be generated by a Liouvillian generator of the form introduced here.
Physically, large pressure perturbations can locally break the simple correspondence
\[
    \text{``hotter} \;\Leftrightarrow\; \text{higher entropy} \;\Leftrightarrow\; \text{lower density''},
\]
so that a fluid particle that is hotter than its surroundings may nevertheless become denser and experience buoyancy braking \citep{Toomre1976,Latour1981}. 
This behaviour is familiar in simulations of compressible convection, where pressure fluctuations and acoustic transients contribute significantly to the force balance in strongly stratified layers \citep{Hurlburt1986,Cattaneo1991}. 
Related limitations arise when convection interacts explicitly with global oscillation modes, for which acoustic propagation and phase lags between pressure, entropy, and velocity perturbations become central ingredients.
Such problems require either time-dependent, non-local convection formalisms coupled to pulsation equations \citep[\eg][]{Gough1977,Xiong1997,Grigahcene2005} or fully compressible radiation-hydrodynamics simulations \citep{Stein2001}.
The present framework is therefore intended primarily for low-Mach-number regimes where pressure remains close to quasi-hydrostatic, and where convection can be treated as a slowly evolving driver of the mean structure; a compressible extension incorporating dynamical pressure fluctuations is left for future work.

\subsection{Phase-space divergence and stability across stellar layers}
\label{sec:discussion_stellar_layers}

The phase-space divergence \(\{s,T\}\) derived in Sect.~\ref{sec:phase_space_divergence} should be read as a local Lagrangian stability criterion, \ie as the instantaneous (de)focusing rate of neighbouring fluid-particle trajectories in \((\vec r,\vec u)\).
A purely positional, layer-wise classification is recovered only in the local quasi-adiabatic limit, where \(\mathrm{sign}\,\{s,T\}=\mathrm{sign}(-N^2)\) (Appendix~\ref{app:schwarzschild_limit}).
Away from that limit, \(\{s,T\}\) remains intrinsically phase-space dependent and must be interpreted in a velocity-resolved manner.

This reduction to a layer-wise stability label is expected to be most useful in deep stellar regions, where locality, quasi-adiabaticity, and small entropy fluctuations are most likely to hold over the bulk of the realised velocity distribution.
In deep radiative layers, fluid-particle excursions over the relevant acceleration time typically remain local with respect to the background stratification, while radiative adjustment acts on thermal timescales that are long compared with the dynamical response.
Entropy fluctuations within the local velocity population therefore remain modest, and the phase-space criterion tends to recover a predominantly negative \(\{s,T\}\), consistently with \(N^2>0\).
In well-developed convective interiors away from boundaries, motions are likewise often close to adiabatic over dynamical times, and the bulk of the realised velocity distribution samples a background stratification that varies only weakly over a typical excursion.
Under such conditions, \(\{s,T\}\) again tends to have a uniform sign at fixed position, now predominantly positive and consistent with \(N^2<0\).
In both cases, the phase-space criterion remains defined in terms of trajectories in \((\vec r,\vec u)\), even though for most realised motions it can still be read as the local counterpart of the classical stable/unstable distinction.

The situation is less straightforward in transition layers, where at least one of the assumptions underlying the quasi-adiabatic reduction breaks down.
Near the stellar surface, radiative exchange may become efficient over dynamical times, so that different parts of the local velocity distribution can experience different entropy evolution, especially between rising and descending motions \citep[\eg][]{Stein1998, Asplund2000a, Asplund2005, Nordlund2009, Freytag2012, Magic2013}.
At convective boundaries, including penetration and overshooting regions, the reduction to a purely positional criterion can also fail because trajectories sample a rapidly changing background stratification over their excursion, so that locality breaks down even when the flow remains nearly adiabatic \citep[\eg][]{Zahn1991, Roxburgh1989, Brummell2002, Rogers2005, Rieutord2019, Korre2019}.
In such regions, \(\{s,T\}\) should therefore be read as an intrinsically phase-space diagnostic rather than as a single label attached to the layer itself.

In practice, while the present derivation isolates the thermal contribution in an inertial, single-fluid setting for clarity, the stability of such layers does not in general reduce to the thermal channel alone.
In a uniformly rotating frame and in a magnetised multi-species plasma, the same strategy still applies, but the effective phase-space brackets acquire additional kinematic (Coriolis/centrifugal) and Lorentz contributions.
Composition effects, in turn, introduce additional non-conservative channels at the species level.
The corresponding derivations, including rotation and magnetism in a multi-species fluid, are given in Appendices~\ref{app:rotation} and \ref{app:multispecies}.

\subsection{Practical use and companion paper}

The present paper is deliberately focused on the structural backbone of phase-space averaging. 
We introduce the fluid-particle distribution \(\rho(t,\vec r,\vec u)\), show that Reynolds-Favre mean-field equations follow as velocity-space moments of a single phase-space continuity law.
For stellar modelling, the remaining practical step is to specify a tractable representation of \(\rho\) from which the velocity-space moments and the resulting transport terms appearing in the mean equations can be computed. 

The phase-space viewpoint suggests a natural modelling route in convective interiors. 
Where the dynamics is predominantly expansive over the velocity range that carries most of the mass at fixed \(\vec r\) (i.e. where \(\rho(t,\vec r,\vec u)\) is appreciable) so that \(\{s,T\}>0\) for most relevant trajectories, Liouvillian trajectories locally diverge and the velocity distribution tends to spread. 
This motivates describing a statistically stationary convective state by a distribution that is maximally mixed in velocity space, subject to the constraints imposed by the quasi-1D stellar background and by conservation laws. 
In transition layers, where the diagnostic becomes strongly velocity-selective, \(\{s,T\}\) provides guidance on where and how departures from a single layer-wise maximum-entropy closure should be expected. 

The companion paper turns this programme into an explicit 1D convection model for stellar envelopes.
Restricting to the reduced phase space \((r,u)\) in stationary, spherically symmetric configurations, it constructs maximum-entropy representations of \(\rho(r,u)\) under the local constraints imposed by the mean stellar structure.
The resulting moments provide the turbulent pressure and energy fluxes entering the one-dimensional structure equations.
Paper~I should therefore be read as the geometric and dynamical backbone of that construction: it defines the averaging operator, the origin of the mean-field hierarchy, and the role of \(\{s,T\}\) in identifying where a mixing-based closure is expected to be appropriate.

\section{Conclusions}
\label{sec:conclusion}

We have introduced a phase-space averaging (PSA) formalism for stellar convection, in which the dynamics of a population of fluid particles is described by a mass distribution \(\rho(t,\vec r,\vec u)\) in position--velocity space rather than by an externally prescribed averaging procedure.
Within this framework, Reynolds-Favre mean-field equations follow exactly as velocity-space moments of a single phase-space continuity law, so that the averaging operator and the associated moment hierarchy are fixed by construction.
Under standard interior conditions, the fluid-particle dynamics can also be written in Liouvillian form through a local generator in \((\vec r,\vec u)\).

A key outcome is the phase-space divergence \(\operatorname{div}(\dot{\vec \varphi})=\{s,T\}\), which measures the instantaneous logarithmic rate of phase-space volume change along the Liouvillian flow.
In the local quasi-adiabatic regime, it recovers the usual layer-wise sign test, \(\mathrm{sign}\,\{s,T\}=\mathrm{sign}(-N^2)\).
In transition layers, however, it remains intrinsically velocity-resolved and should be interpreted as a Lagrangian stability criterion in phase space.

The companion paper exploits this phase-space viewpoint to build practical closures in one-dimensional envelopes by constructing tractable representations of \(\rho(r,u)\) and evaluating the transport terms that enter the mean equations from its moments.
More broadly, the framework makes explicit how additional physics can be incorporated at the phase-space level, through geometric modifications of the brackets and through extra non-conservative channels when new degrees of freedom are introduced.
Taken together, Paper~I and Paper~II aim to provide a ``Liouvillian route'' to stellar convection modelling in which averaged equations, stability criteria, and closure assumptions stem from a single, dynamically grounded phase-space picture.

\begin{acknowledgements}
This work was funded by the European Research Council (ERC) under the Horizon Europe programme (Synergy Grant agreement No. 101071505: 4D-STAR). 
While partially funded by the European Union, views and opinions expressed are however those of the authors only and do not necessarily reflect those of the European Union or the European Research Council. 
Neither the European Union nor the granting authority can be held responsible for them.
\end{acknowledgements}

\bibliographystyle{aa}
\bibliography{src}

\appendix
\nolinenumbers

\section{Local quasi-adiabatic limit: sign-consistency with Schwarzschild}
\label{app:schwarzschild_limit}

In this appendix, we derive the sign-consistency relation
\[
    \mathrm{sign}\,\{s,T\}=\mathrm{sign}(-N^2),
\]
used in Sect.~\ref{sec:phase_space_divergence}.
Throughout, we retain the assumptions of Sect.~\ref{sec:liouvillian_generator}, namely immediate pressure adjustment and negligible viscous stresses.

\subsection{From phase-space divergence to a buoyancy form}
\label{app:schwarzschild_limit_buoyancy}

We begin from the identity \(\operatorname{div} \, (\dot{\vec \varphi})=\{s,T\}\).
Using in addition the standard thermodynamic Jacobian identity
\[
    \left| \dfrac{\partial (s, T)}{\partial (v, p)}\right| = 1,
\]
we may rewrite the bracket as
\begin{equation}
    \label{eq:app_divphi_buoyancy}
    \operatorname{div} \, (\dot{\vec \varphi})
    = \left| \dfrac{\partial (s, T)}{\partial (v, p)}\right| \{v, p\}
    = \{v, p\}
    = - \grad_{u} v \cdot \grad_{r} p,
\end{equation}
since \(\grad_{u} p=\vec 0\).
In this form, the diagnostic naturally appears as a measure of the velocity-sensitivity of buoyancy.

To make this more explicit, we now relate \(\grad_{u} v\) to \(\grad_{u} s\).
At fixed pressure,
\begin{equation}
    \label{eq:app_gradv_to_grads}
    \grad_{u} v
    = \left( \dfrac{\partial v}{\partial s} \right)_p \grad_{u} s
    = \frac{\nu_p\,v}{c_p}\,\grad_{u} s,
\end{equation}
where \(\nu_p\equiv(\partial\ln v/\partial\ln T)_p\) is the thermal dilatation coefficient.
Substituting this into Eq.~\eqref{eq:app_divphi_buoyancy} then yields
\begin{equation}
    \label{eq:app_bracket_explicit_grads}
    \{s,T\}
    = -\left(\frac{\nu_p}{c_p}\,\grad_{u} s\right)\cdot\big(v\,\grad_{r} p\big).
\end{equation}

At the same time, the Brunt-V\"ais\"al\"a frequency may be written in the corresponding entropy form as
\begin{equation}
    \label{eq:app_N2_entropy_form}
    -\,N^2
    = \left(\frac{\nu_p}{c_p}\,\grad_{r}\tilde s\right)\cdot\big(\tilde v\,\grad_{r} p\big).
\end{equation}
The two expressions now differ only through the replacement of \(\grad_{r}\tilde s\) by \(-\,\grad_{u} s\), together with the distinction between \(v\) and \(\tilde v\).
To establish sign-consistency, it is therefore enough to show that, in the local quasi-adiabatic regime, \(\grad_{u} s\) is approximately antiparallel to \(\grad_{r}\tilde s\) over the relevant part of velocity space.

\subsection{Local quasi-adiabatic approximation}
\label{app:schwarzschild_limit_tau}

We now derive a local trajectory-based approximation for \(\grad_{u} s\).
More precisely, we show that
\[
    \grad_{u} s \simeq -\,\tau\,\grad_{r} \tilde s.
\]

\begin{enumerate}
\item[(i)] \textit{Locality and acceleration-time anchoring.}
Consider a given phase-space point \((t,\vec r,\vec u)\), and let \(\tau>0\) be the backward time needed for the Liouvillian trajectory to reach \(\vec u=\vec 0\), so that
\(\vec u(t-\tau)=\vec 0\).
Over such a short interval, the corresponding displacement is, to first order,
\begin{equation}
    \label{eq:app_local_displacement}
    \vec r_0 \equiv \vec r(t-\tau) \simeq \vec r - \tau\,\vec u,
\end{equation}
which simply expresses the requirement that the excursion remain local relative to the background stratification.

\item[(ii)] \textit{Quasi-adiabaticity over \(\tau\).}
Along the same trajectory, entropy satisfies
\begin{equation}
    \label{eq:app_entropy_char}
    s(t,\vec r,\vec u)
    = s(t-\tau,\vec r_0,\vec 0)
      + \int_{t-\tau}^{t}\dot s\big(t',\vec\varphi(t')\big)\,dt',
\end{equation}
where \(\dot s\) denotes the entropy derivative along the trajectory.
If the motion is locally quasi-adiabatic over the interval \(\tau\), then the non-advective entropy exchange remains subdominant.
To leading order, we may therefore write
\begin{equation}
    \label{eq:app_quasi_adiabatic_drop}
    s(t,\vec r,\vec u)\simeq s(t-\tau,\vec r_0,\vec 0).
\end{equation}

\item[(iii)] \textit{Small entropy fluctuations within the local velocity population.}
We next write \(s=\tilde s+s''\) at fixed \((t,\vec r)\), and assume that entropy fluctuations remain small over the relevant local velocity population.
In particular, at the anchoring point \(\vec u=\vec 0\), the entropy is close to the local Favre mean, so that
\begin{equation}
    \label{eq:app_anchor_to_mean}
    s(t-\tau,\vec r_0,\vec 0)
    = \tilde s(t-\tau,\vec r_0) + s''(t-\tau,\vec r_0,\vec 0)
    \simeq \tilde s(t-\tau,\vec r_0).
\end{equation}
\end{enumerate}

Taken together, Eqs.~\eqref{eq:app_quasi_adiabatic_drop} and \eqref{eq:app_anchor_to_mean} lead to the central approximation
\begin{equation}
    \label{eq:app_s_to_mean}
    s(t,\vec r,\vec u)\simeq \tilde s(t-\tau,\vec r_0).
\end{equation}
We may then expand the right-hand side about \((t,\vec r)\) using Eq.~\eqref{eq:app_local_displacement}.
To first order, this gives
\begin{equation}
    \label{eq:app_taylor_mean}
    \tilde s(t-\tau,\vec r_0)
    \simeq \tilde s(t,\vec r)
      - \tau\,\partial_t\tilde s(t,\vec r)
      - \tau\,\vec u\cdot\grad_{r}\tilde s(t,\vec r).
\end{equation}
Differentiating Eq.~\eqref{eq:app_s_to_mean} with respect to \(\vec u\) at fixed \((t,\vec r)\), and retaining only the leading-order contribution, we finally obtain
\begin{equation}
    \label{eq:app_gradu_s_first_order}
    \grad_{u} s(t,\vec r,\vec u)
    \simeq -\,\tau\,\grad_{r}\tilde s(t,\vec r),
\end{equation}
with \(\tau>0\) by construction.
The neglected corrections arise from the entropy-source integral in \eqref{eq:app_entropy_char}, from higher-order non-locality in \(\vec r_0\), and from finite entropy fluctuations within the local velocity population.

\subsection{Sign-consistency with Schwarzschild}
\label{app:schwarzschild_limit_sign}

With this result in hand, we return to Eq.~\eqref{eq:app_bracket_explicit_grads}.
Substituting Eq.~\eqref{eq:app_gradu_s_first_order} into it gives, at leading order,
\begin{equation}
    \label{eq:app_bracket_to_N2}
    \{s,T\}
    \simeq
    \tau\,\left(\frac{\nu_p}{c_p}\,\grad_{r}\tilde s\right)\cdot\big(v\,\grad_{r} p\big).
\end{equation}
As entropy fluctuations remain small, \(v\simeq \tilde v\).
Comparing with Eq.~\eqref{eq:app_N2_entropy_form}, we therefore recover
\begin{equation}
    \label{eq:app_sign_match_final}
    \mathrm{sign}\,\{s,T\} = \mathrm{sign}\,(-N^2),
\end{equation}
since \(\tau>0\) and \(\nu_p/c_p>0\).

Conversely, outside the local quasi-adiabatic regime, the terms neglected in
\eqref{eq:app_entropy_char}--\eqref{eq:app_taylor_mean} restore an explicit velocity dependence through both non-local excursions and \(\dot s\).
In that case, \(\{s,T\}\) remains intrinsically phase-space dependent.

\section{Liouvillian structure in the presence of rotation}
\label{app:rotation}

This appendix illustrates how the Liouvillian structure is modified when the phase-space variables are expressed in a uniformly rotating frame.
We consider the change of variables from an inertial frame to a uniformly rotating frame, \((\vec{x}, \dot{\vec{x}}) \to (\vec r, \vec{c})\), defined by
\begin{equation}
    \vec{x} = \vec{x}_0 + R(t) \cdot \vec r,
\end{equation}
where \(R(t)\) is an orthogonal rotation matrix and \(\vec{x}_0\) is the origin of the rotating frame.

\subsection{Liouvillian in a rotating frame}

A convenient generating function to construct the transformed Liouvillian is
\begin{equation}
    f(\vec x,\vec c) = \vec r \cdot \vec{c},
    \qquad
    \vec r = R^\dag(t)\cdot(\vec x - \vec x_0),
\end{equation}
which yields the relation between inertial and canonical velocities,
\begin{equation}
    \dot{\vec{x}} = \grad_{\vec{x}} f = R \cdot \vec{c},
\end{equation}
that is, the canonical velocity \(\vec c\) equals the inertial velocity expressed in the rotating basis.
The transformed Liouvillian reads
\begin{equation}
\begin{split}
    \Psi
    &= \psi + \partial_t f \\
    &= \psi + (\vec{x} - \vec{x}_0) \cdot \dot{R} \cdot \vec{c} \\
    &= \psi + (\vec{x} - \vec{x}_0) \cdot R \cdot (R^\dag \cdot \dot{R}) \cdot \vec{c} \\
    &= \psi - \vec{c} \cdot \Omega \cdot \vec r,
\end{split}
\end{equation}
with
\begin{equation}
    \Omega = R^\dag \cdot \dot{R}
\end{equation}
the antisymmetric angular-velocity tensor associated with \(R(t)\).
Introducing the kinematic velocity \(\vec u\) as
\begin{equation}
    \vec u \equiv \dot{\vec r} = \grad_{\vec{c}} \Psi,
\end{equation}
one finds the relation between kinematic and canonical velocities,
\begin{equation}
    \vec u = \vec{c} - \Omega \cdot \vec r,
    \qquad \Rightarrow \qquad
    \vec{c} = \vec u + \Omega \cdot \vec r.
\end{equation}

For physical interpretation it is often preferable to work with the kinematic velocity \(\vec u\) rather than the canonical one \(\vec{c}\).
Expressing \(\Psi\) in the variables \((\vec r, \vec u)\) gives
\begin{equation}
\begin{split}
    \Psi(\vec r,\vec u)
    &= \tfrac12\,\vec u^{\,2}
       - \tfrac12\,(\Omega \cdot \vec r)^2
       + \phi(\vec r)
       + \int v\, dp,
\end{split}
\end{equation}
which exhibits the centrifugal potential contribution \(- \tfrac12\,(\Omega \cdot \vec r)^2\) on top of gravity and pressure work.

\subsection{Equations of motion}

Working with the non-canonical variables \((\vec r, \vec u)\) twists the symplectic two-form \(\omega\), which now depends on the angular-velocity tensor \(\Omega\),
\begin{equation}
\begin{split}
    \omega
    &= \1 : (d\vec r \wedge d\vec{c}) \\
    &= \1 : \big(d\vec r \wedge d(\vec u + \Omega\cdot\vec r)\big) \\
    &= \1 : (d\vec r \wedge d\vec u)
       -\,\Omega : (d\vec r \wedge d\vec r).
\end{split}
\end{equation}

Let
\begin{equation}
    X_\Psi = \dot{\vec r} \cdot \grad_{r} + \dot{\vec u} \cdot \grad_{u}
\end{equation}
denote the Liouvillian vector field on \((\vec r, \vec u)\).
It satisfies
\begin{equation}
    \label{eq:Liouvillian_differential_RHS}
    X_\Psi \cdot \omega = d\Psi
    = \grad_{r} \Psi \cdot d\vec r + \grad_{u} \Psi \cdot d\vec u.
\end{equation}
Using the identity \(\vec{a} \cdot (\vec{b} \wedge \vec{c}) = (\vec{a}\cdot\vec{b})\,\vec{c} - (\vec{a}\cdot\vec{c})\,\vec{b}\) to expand the left-hand side of Eq.~\eqref{eq:Liouvillian_differential_RHS}, one obtains
\begin{equation}
\begin{split}
    \label{eq:Liouvillian_differential_LHS}
    X_\Psi \cdot \omega
    &= -\,\Omega : (\dot{\vec r}\, d\vec r)
       + \Omega^\dag : (\dot{\vec r}\, d\vec r)
       + \1 : (\dot{\vec r}\, d\vec u)
       - \1 : (\dot{\vec u}\, d\vec r) \\
    &= -\,(\dot{\vec u} + 2 \Omega \cdot \vec u) \cdot d\vec r
       + \dot{\vec r} \cdot d\vec u.
\end{split}
\end{equation}
Term-by-term identification between Eqs.~\eqref{eq:Liouvillian_differential_RHS} and \eqref{eq:Liouvillian_differential_LHS} gives the equations of motion,
\begin{align}
    \dot{\vec r} &= \grad_{u} \Psi = \vec u, \\
    \dot{\vec u} &= -\,\grad_{r} \Psi - 2 \Omega \cdot \vec u
                 = \vec{g} - v\,\grad_{r} p - \Omega^2 \cdot \vec r - 2 \Omega \cdot \vec u,
\end{align}
namely the usual centrifugal acceleration \(-\,\Omega^2 \cdot \vec r\) and Coriolis acceleration \(-\,2 \Omega \cdot \vec u\) in addition to gravity and pressure forces.

\subsection{Poisson brackets and phase-space divergence}

Poisson brackets follow from inverting the twisted symplectic form and inherit the rotational torsion. 
For scalar observables \(f(\vec r,\vec u)\) and \(g(\vec r,\vec u)\) we denote the resulting bracket by \(\{\cdot,\cdot\}_\Omega\), with elementary relations
\begin{equation*}
    \begin{alignedat}{2}
        \{\vec r, \vec r\}_\Omega &= \0,        &\qquad \{\vec r, \vec u\}_\Omega &= \1,\\
        \{\vec u, \vec r\}_\Omega &= -\,\1,     &\qquad \{\vec u, \vec u\}_\Omega &= -\,\Omega + \Omega^\dag = -\,2 \Omega .
    \end{alignedat}
\end{equation*}
Equivalently,
\begin{equation}
    \{f,g\}_\Omega
    = \grad_{r} f \cdot \grad_{u} g
      - \grad_{u} f \cdot \grad_{r} g
      - 2 \vec{\Omega} \cdot \big(\grad_{u} f \times \grad_{u} g\big),
\end{equation}
where \(\vec\Omega\) is the rotation vector defined by \(\Omega\cdot\vec a=\vec\Omega\times\vec a\).

Uniform rotation does not introduce an additional non-conservative contribution to the Liouvillian differential, but it does modify the phase-space divergence through the torsion of the bracket. 
In particular, the thermal contribution to the phase-space divergence can be written as
\begin{equation}
    \operatorname{div} \, (\dot{\vec \varphi}) = \{s,T\}_\Omega,
\end{equation}
so Coriolis effects enter through
\begin{equation}
    \{s, T\}_\Omega
    = \grad_{r} s \cdot \grad_{u} T
      - \grad_{u} s \cdot \grad_{r} T
      - 2 \vec{\Omega} \cdot \big(\grad_{u} s \times \grad_{u} T\big).
\end{equation}

\section{Multi-species Liouvillian structure: Lorentz force and compositional effects}
\label{app:multispecies}

This appendix generalises the Liouvillian formalism to a magnetised multi-species fluid.
As in the rotating case, magnetic effects enter geometrically through a twisted phase-space bracket, whereas compositional degrees of freedom contribute through additional non-conservative terms.

\subsection{Multi-species phase-space description}

We start from a species-resolved phase-space description.
For each species \(s\), with mass \(m_s\) and charge \(q_s\), we denote by
\[
    \rho_s(t,\vec r,\vec u)
\]
the mass density of species-\(s\) fluid particles in phase space.
It is convenient to introduce the corresponding charge density in phase space,
\begin{equation}
    \sigma_s(t,\vec r,\vec u)
    = \alpha_s\,\rho_s(t,\vec r,\vec u),
    \qquad
    \alpha_s \equiv q_s / m_s,
\end{equation}
which mirrors the usual kinetic-theory relation between particle and charge distributions.

The total phase-space mass density is
\begin{equation}
    \rho(t,\vec r,\vec u) = \sum_s \rho_s(t,\vec r,\vec u),
\end{equation}
and the mass fraction of species \(s\) attached to a given fluid particle is
\begin{equation}
    X_s(t,\vec r,\vec u) = \frac{\rho_s}{\rho},
    \qquad
    \sum_s X_s = 1.
\end{equation}
Accounting for composition modifies the equation of state at the fluid-particle level, and the specific internal energy of species \(s\) admits the differential
\begin{equation}
    \label{eq:differential_eps_s}
    d \varepsilon_s
    = T ds_s
      - p dv_s
      + \mu_s\,d X_s,
\end{equation}
where \(v_s\) and \(\mu_s\) denote respectively the specific volume and chemical potential carried by a species-\(s\) fluid particle.
This structure will be reflected in the Liouvillian for each species through additional terms in \(d\psi_s\), on the same footing as entropy.

Averaged quantities for species \(s\) are obtained as velocity-space moments of \(\rho_s\) and \(\sigma_s\), namely
\begin{align}
    \label{eq:density_s}
    \bar\rho_s(\vec r,t)
    &= \int \rho_s(t,\vec r,\vec u)\, d\vec u, \\
    \label{eq:charge_density_s}
    \bar\sigma_s(\vec r,t)
    &= \int \sigma_s(t,\vec r,\vec u)\, d\vec u
     = \alpha_s\,\bar\rho_s, \\
    \label{eq:charge_current_density_s}
    \vec j_s(\vec r,t)
    &= \int \sigma_s(t,\vec r,\vec u)\,\vec u\, d\vec u.
\end{align}
The total mass and charge currents are then
\begin{equation}
    \label{eq:total_density}
    \bar\rho = \sum_s \bar\rho_s,
    \qquad
    \bar\sigma = \sum_s \bar\sigma_s,
    \qquad
    \vec j = \sum_s \vec j_s,
\end{equation}
which allows us to recover the standard composition fractions as the Favre averages of \(X_s\),
\begin{equation}
    \tilde X_s(\vec r,t)
    = \frac{\langle \rho X_s \rangle}{\bar\rho}
    = \frac{\bar\rho_s}{\bar\rho}, 
    \qquad \sum_s \tilde{X}_s = 1.
\end{equation}

In the stellar context we are chiefly interested in quasi-neutral plasmas, for which the net charge density vanishes to leading order,
\begin{equation}
    \bar\sigma(\vec r,t) = \sum_s \bar\sigma_s(\vec r,t) \simeq 0.
\end{equation}
Quasi-neutrality does not, however, imply that the total current vanishes: \(\vec j\) remains finite and is responsible for the macroscopic Lorentz force in the one-fluid limit.

\subsection{Magnetic torsion in phase space for a single species}
\label{app:multispecies_magnetic_torsion}

In the MHD regime considered here, charge separation is negligible and the electric field can be taken to vanish in the comoving frame,
\begin{equation}
    \vec E \simeq \vec 0,
\end{equation}
so that electromagnetic forces reduce to the magnetic part of the Lorentz force.
In the Liouvillian picture, this contribution can be incorporated geometrically, in direct analogy with the rotational torsion of Appendix~\ref{app:rotation}, by working with a vector potential \(\vec A(\vec r)\) such that
\begin{equation}
    \vec B = \grad \times \vec A.
\end{equation}

For a given species \(s\), we introduce a canonical velocity \(\vec c_s\) related to the kinematic velocity \(\vec u\) by
\begin{equation}
    \label{eq:canonical_u_relation}
    \vec c_s = \vec u + \alpha_s \vec A(\vec r).
\end{equation}
We regard \((\vec r,\vec c_s)\) as canonical phase-space coordinates for species \(s\), equipped with the standard symplectic form
\[
    \omega_s = \1 : (d\vec r \wedge d\vec c_s).
\]
Expressed in the non-canonical variables \((\vec r,\vec u)\), the symplectic form becomes
\begin{equation}
\begin{split}
    \omega_s
    &= \1 : \big(d\vec r \wedge d\vec u\big)
       + \alpha_s\,\1 : \big(d\vec r \wedge d\vec A\big) \\
    &= \1 : \big(d\vec r \wedge d\vec u\big)
       + \alpha_s\,\grad \vec{A} : \big(d\vec r \wedge d\vec r\big) \\
    &= \1 : \big(d\vec r \wedge d\vec u\big)
       - \frac{\alpha_s}{2} \,  {B} : \big(d\vec r \wedge d\vec r\big),
\end{split}
\end{equation}
where the antisymmetric tensor \( {B}\) is defined by
\begin{equation}
     {B} = - \grad \wedge \vec A,
    \qquad \text{i.e.}\quad
    (  B)_{ij} = - \epsilon_{ijk} B_k,
    \qquad
     {B}\cdot\vec a = \vec B \times \vec a.
\end{equation}
The magnetic field thus twists the symplectic two-form on \((\vec r,\vec u)\), exactly as the angular-velocity tensor does in the rotating case (Appendix~\ref{app:rotation}), but now through the vector potential.

\subsection{Liouvillian and equations of motion for a single species}

We now consider the Liouvillian vector field of species \(s\) on \((\vec r, \vec u)\),
\begin{equation}
    X_{\psi_s} = \dot{\vec r} \cdot \grad_{r}
               + \dot{\vec u} \cdot \grad_{u}.
\end{equation}

In the presence of compositional degrees of freedom, the thermodynamic part of the Liouvillian of species \(s\) acquires the differential
\begin{equation}
\begin{split}
    \label{eq:differential_psi_s}
    d\psi_s
    &= d\Big(
        \tfrac12 \vec u^2 + \varepsilon_s + p v_s + \phi
    \Big)
      - T ds_s
      - \mu_s\,dX_s \\
    &= d\Big(
        \tfrac12 \vec u^2 + \phi
    \Big)
      + v_s\,dp,
\end{split}
\end{equation}
where we have expanded \(d\varepsilon_s\) using Eq.~\eqref{eq:differential_eps_s}, so that all terms involving \(ds_s\), \(dv_s\) and \(dX_s\) cancel.

The equations of motion follow, as in Appendix~\ref{app:rotation}, from the defining relation
\begin{equation}
    X_{\psi_s} \cdot \omega_s = d\psi_s
    =   \grad_{r} \psi_s \cdot d\vec r
      + \grad_{u} \psi_s \cdot d\vec u,
\end{equation}
which yields
\begin{equation}
\begin{split}
    X_{\psi_s} \cdot \omega_s
    = -\,\big(\dot{\vec u}
               - \alpha_s\,\dot{\vec r}\times\vec B\big)\cdot d\vec r
       + \dot{\vec r}\cdot d\vec u.
\end{split}
\end{equation}
Term-by-term identification with \(d\psi_s\) then gives
\begin{align}
    \dot{\vec r} &= \grad_{u} \psi_s = \vec u, \\
    \label{eq:momentum_mag_s}
    \dot{\vec u} &= - \grad_{r} \psi_s
                    + \alpha_s\,\dot{\vec r} \times \vec B
                 = \vec{g} - v_s\,\grad p
                    + \frac{q_s}{m_s}\,\vec u\times\vec B,
\end{align}
namely the usual Lorentz force for species \(s\) in the magnetically dominated regime.

\subsection{Species-wise phase-space diagnostic}
\label{app:multispecies_stability}

The twisted symplectic form \(\omega_s\) associated with species \(s\) induces a Poisson bracket \(\{\cdot,\cdot\}_s\) on \((\vec r,\vec u)\), which reduces to the bracket of Sect.~\ref{sec:liouvillian_generator} in the non-magnetic case and inherits a magnetic torsion proportional to \(\alpha_s \vec B\).
For scalar observables \(f(t,\vec r,\vec u)\) and \(g(t,\vec r,\vec u)\) one may write
\begin{equation}
    \{f,g\}_s
    = \grad_{r} f \cdot \grad_{u} g
      - \grad_{u} f \cdot \grad_{r} g
      - \alpha_s\,\vec{B} \cdot \big(\grad_{u} f \times \grad_{u} g\big),
\end{equation}
which makes explicit the magnetic contribution through the last term.

In the single-species, non-compositional case, the non-conservative contribution \(-T ds\) in \(d\psi\) leads to an entropy balance in phase space of the form
\begin{equation}
    \operatorname{div} \, (\dot{\vec \varphi})
    = \Dt (-\ln\rho)
    = \{s, T\},
\end{equation}
as shown in Sect.~\ref{sec:entropy_balance}.
For species \(s\) in the present multi-species setting, the defining decomposition
\[
    d\psi_s = dH_s - T\,ds_s - \mu_s\,dX_s
\]
shows that compositional degrees of freedom enter on the same footing as entropy at the fluid-particle level.
Although these non-conservative differentials cancel from the simplified expression of \(d\psi_s\) once the first law is used (as in the single-species case), they control the non-conservative part of the phase-space divergence and therefore contribute explicitly to the divergence-based stability balance.
The phase-space continuity equation for \(\rho_s\) can then be rewritten as
\begin{equation}
    \label{eq:div_species_full}
    \operatorname{div}\,(\dot{\vec\varphi}_s)
    = \Dt (-\ln\rho_s)
    = \{s_s, T\}_s + \{X_s, \mu_s\}_s,
\end{equation}
where \(\vec\varphi_s=(\vec r,\vec u)\) and the two brackets on the right-hand side represent, respectively, the thermal and compositional channels of non-conservative evolution for species \(s\).
The first term,
\begin{equation}
    \{s_s, T\}_s
    = \grad_{r} s_s \cdot \grad_{u} T
      - \grad_{u} s_s \cdot \grad_{r} T
      - \alpha_s\,\vec{B} \cdot \big(\grad_{u} s_s \times \grad_{u} T\big),
\end{equation}
is the direct analogue of the diagnostic introduced in Sect.~\ref{sec:entropy_balance}, but now evaluated with the species-wise bracket \(\{\cdot,\cdot\}_s\) including magnetic torsion.
The second term,
\begin{equation}
    \{X_s, \mu_s\}_s
    = \grad_{r} X_s \cdot \grad_{u} \mu_s
      - \grad_{u} X_s \cdot \grad_{r} \mu_s
      - \alpha_s\,\vec{B} \cdot \big(\grad_{u} X_s \times \grad_{u} \mu_s\big),
\end{equation}
collects the compositional contribution and vanishes when \(X_s\) is materially conserved and \(\mu_s\) is spatially and kinematically uniform.

Equation~\eqref{eq:div_species_full} shows that, for each species \(s\), the phase-space diagnostic is controlled by the combined action of thermal and compositional gradients, both filtered through the magnetically twisted bracket.
A negative right-hand side corresponds to phase-space contraction for species \(s\) (stabilising behaviour), whereas a positive right-hand side signals phase-space expansion and hence a destabilising tendency for that species.
The decomposition into \(\{s_s,T\}_s\) and \(\{X_s,\mu_s\}_s\) identifies, respectively, the purely thermal and compositional contributions to this tendency.

A global stability criterion for the multi-species mixture would in principle require analysing the coupled set of variables \(\{s_s,X_s\}_s\) and their mutual interactions (collisions, diffusion, drift motions), and constructing an appropriate mixture-wide diagnostic from the individual brackets in Eq.~\eqref{eq:div_species_full}.
A detailed mixture-level analysis, which would provide the phase-space analogue of the classical Ledoux criterion, is left for future investigation.

\subsection{One-fluid limit and emergence of the macroscopic Lorentz force}
\label{app:multispecies_sum}

For each species \(s\), mass conservation in phase space is expressed by
\begin{equation}
    \partial_t \rho_s
    + \grad_{r} \cdot(\rho_s \vec u)
    + \grad_{u} \cdot(\rho_s \vec\gamma_s) = 0,
\end{equation}
with
\begin{equation}
    \vec\gamma_s
    = \vec g - v_s\,\grad p
      + \alpha_s\,\vec u\times\vec B
\end{equation}
resulting from Eq.~\eqref{eq:momentum_mag_s}.
Applying the general moment procedure of Sect.~\ref{sec:psa} to this continuity equation, with \(x_s=\vec u\), yields the momentum equation for species \(s\),
\begin{equation}
    \partial_t\langle \rho_s \vec u \rangle
    + \grad \cdot \Big(
        \langle \rho_s \vec u \rangle \,\tilde{\vec u}_s
        + \mathrm R_s
      \Big)
    = \langle \rho_s \vec\gamma_s \rangle,
\end{equation}
where the velocity-space average is
\begin{equation}
    \langle \bullet \rangle
    = \int \bullet \, d\vec u,
\end{equation}
and
\begin{equation}
    \tilde{\vec u}_s
      = \frac{\langle \rho_s \vec u \rangle}{\bar\rho_s},
    \qquad
    \mathrm R_s = \langle \rho_s \vec u_s''\vec u_s''\rangle,
    \qquad
    \vec u_s'' = \vec u - \tilde{\vec u}_s,
\end{equation}
are the Favre-mean velocity and Reynolds tensor of species \(s\).
Using the explicit form of \(\vec\gamma_s\) and the identities
\eqref{eq:density_s}-\eqref{eq:charge_current_density_s}, one obtains
\begin{equation}
    \partial_t\langle \rho_s \vec u \rangle
    + \grad \cdot \Big(
        \langle \rho_s \vec u \rangle \,\tilde{\vec u}_s
        + \mathrm R_s
      \Big)
    = \bar\rho_s \vec g
      - \tilde{X}_s \grad p
      + \vec j_s \times \vec B .
\end{equation}
Thus, for each species, the magnetic contribution to the momentum balance takes the form of a Lorentz force density \(\vec j_s\times\vec B\), with the current \(\vec j_s\) defined as the phase-space moment of the charge-weighted velocity.

The usual one-fluid description is recovered by summing these equations over all species.
Introducing the total mass density and current as in Eq.~\eqref{eq:total_density} and the total Reynolds tensor
\[
    \mathrm R = \sum_s \mathrm R_s,
\]
one finds
\begin{equation}
\begin{split}
    \partial_t \sum_s\langle \rho_s \vec u \rangle
    + \grad \cdot \Big(
        \sum_s \langle \rho_s \vec u \rangle \,\tilde{\vec u}_s
        + \mathrm R
      \Big) 
    = \bar\rho\,\vec g
       - \grad p
       + \vec j \times \vec B .
\end{split}
\end{equation}
Because summation and velocity integration commute,
\begin{equation}
    \sum_s \langle \rho_s \vec u \rangle = \Big\langle (\sum_s \rho_s) \vec u \Big\rangle = \langle \rho \vec u \rangle,
\end{equation}
the left-hand side may be expressed in terms of the one-fluid Favre velocity
\begin{equation}
\begin{split}
    \partial_t \langle \rho \vec u \rangle
    + \grad \cdot \Big(
        \sum_s \langle \rho_s \vec u \rangle \,\tilde{\vec u}_s
        + \mathrm R
      \Big) 
    = \bar\rho\,\vec g
       - \grad p
       + \vec j \times \vec B .
\end{split}
\end{equation}
In general, differences between \(\tilde{\vec u}_s\) and the one-fluid Favre velocity \(\tilde{\vec u}\) encode inter-species drift and diffusive transport in compositionally stratified regions.
In the strict one-fluid limit where these drifts are neglected, \(\tilde{\vec u}_s \simeq \tilde{\vec u}\) and the flux term reduces to the familiar MHD form, while the macroscopic Lorentz force \(\vec j\times\vec B\) remains the sum of the species-wise Lorentz forces generated by the magnetic torsion.

\end{document}